# Direct Observation of Valley Coupled Topological Current in $MoS_2$

Terry Y.T. Hung[1,2], Kerem Y. Camsari[1], Shengjiao Zhang[1,2], Pramey Upadhyaya[1], and Zhihong Chen[1,2]*

[1]School of Electrical and Computer Engineering &

[2]Birck Nanotechnology Center

Purdue University, West Lafayette, IN 47907

*Correspondence to: zhchen@purdue.edu

**Abstract:** The valley degree of freedom of electrons in two-dimensional transition metal dichalcogenides has been extensively studied by theory (*1–4*), optical (*5–9*) and optoelectronic (*10–13*) experiments. However, generation and detection of pure valley current without relying on optical selection have not yet been demonstrated in these materials. Here, we report that valley current can be electrically induced and detected through the valley Hall effect and inverse valley Hall effect, respectively, in monolayer molybdenum disulfide. Specifically, long-range valley transport is observed over half a micron distance at room temperature. Our findings will enable a new generation of electronic devices utilizing the valley degree of freedom, which can be used for future novel valleytronic applications.

**One Sentence Summary:** We report the first electrical generation and detection of valley currents in monolayer molybdenum disulfide, which is a step forward towards novel information processing and storage through valley degree of freedom of electrons.

**Main Text:**

Electronic devices exploring carrier transport with spin and valley degree of freedom (DOF) have emerged as promising candidates for next-generation information storage and transport, since pure spin and valley currents do not accompany energy dissipation associated with Joule heating. The ability to electrically generate and detect such pure spin and valley currents in these devices is of particular importance. Over the last decade, driven by the emergence of the spin-orbit coupling engineering, tremendous experimental progress has been made to efficiently generate spin currents by electric currents. On the other hand, electrical control of the valley DOF has just started to attract interests in the past few years, initiated by theoretical studies of valleytronics in two dimensional honeycomb lattice systems, such as gapped graphene and transition metal dichalcogenides (TMDs) (*1–4, 14, 15*), revealing the interplay of their unique band structures and topologies. Experimentally, topological valley transport has been observed in graphene systems when a superlattice structure or perpendicular electric field is employed to break the inversion symmetry of this zero bandgap semiconductor (*16–18*).

In contrast, monolayer TMDs, such as molybdenum disulfide ($MoS_2$), is a direct bandgap semiconductor. Electronic transport in these materials is dominated by the inequivalent *K* and *K'* valleys of the Brillouin zone located at band edges. Because of the inherent absence of inversion symmetry in monolayer TMDs, carriers in these two valleys possess non-zero Berry curvature (**Ω**) without needing the assistance of external mechanisms to break the symmetry like in graphene systems. Importantly, *K* and *K'* valleys are related by time-reversal symmetry, which forces Berry curvature to flip its sign, i.e. $\mathbf{\Omega}(K) = -\mathbf{\Omega}(K')$, and allows for optical selection through optical pumping of valley polarization (*5–7*). **Ω** acts as a pseudo-magnetic field in the momentum space and results in an anomalous transverse velocity in the presence of an electric field, i.e. $v_\perp = -\frac{e}{\hbar} E \times \Omega(k)$. Consequently, carriers from $K$ and $K'$ valleys develop opposite $v_\perp$, providing a route to electrically generate pure valley currents transverse to the applied electric field. This so-called valley Hall effect (VHE) has been employed by Mak et al. in monolayer MoS$_2$ devices to measure valley polarization created by circularly polarized light (*10*) and has successfully generated polarization in gated bilayer MoS$_2$ that was then visualized by Kerr rotation microscopy (*11*). It is important to note that this unique VHE phenomenon would not appear in multi-layer MoS$_2$ devices because inversion symmetry holds in multi-layers and carrier transport does not involve K and K' valleys in these indirect bandgap materials, which can be used as a direct comparison or control sample to monolayer devices. Figure 1A illustrates the VHE occurring in the left vertical electrode of a monolayer MoS$_2$ Hall bar device. Analogous to the spin current, such valley current comprises of carriers of opposite (valley) polarization moving along opposite directions, resulting in charge neutral valley current along *x*-axis. Onsager reciprocity (*19*) then ensures the reciprocal effect, a phenomenon defined as the inverse valley Hall effect (iVHE) that converts a non-zero valley current into a transverse electric field, and finally develops charge accumulation across the right vertical electrode of the Hall bar in Fig. 1A. In this paper, we demonstrate electrical generation and detection of valley current in monolayer MoS$_2$ by combining VHE and iVHE in the above-described non-local Hall bar device geometry. We observe large non-local signals at distances more than half micrometer away from the charge current path and a unique temperature dependence that is consistent with valley transport physics.

A colored scanning electron microscopy (SEM) image of one of the MoS$_2$ Hall bar devices measured is shown in Fig. 1C. Two types of measurements can be made, as illustrated in Fig. 1B.

A conventional four probe measurement (type II) allows the extraction of sheet resistance and contact resistance, while the non-local set up (type I) measures the Hall voltage induced by any carrier distributions due to the valley Hall effect or classical Ohmic contribution. A back gate voltage ($V_g$) is applied to the $SiO_2$/Si substrate in order to modulate the carrier concentration in the $MoS_2$ channel. Device fabrication and measurement details are provided in (*20*), section 1.1 and 1.2. Typical n-type $MoS_2$ field-effect transistor behaviors are observed in two probe measurements of all devices; sheet resistance and contact resistances are extracted from type II measurements for various temperatures ranging from 4K to 300K [see (*20*), section 1.2]. Field effect mobility of ~10 $cm^2$/Vs is typically measured for monolayer devices at room temperature.

The most important spurious signal to be ruled out in our measurements is the Ohmic contribution that can result in a van der Pauw like signal (*21*) in a typical non-local, type I measurement. When a DC bias of $V_{ds}$ = 5V is applied to the left electrode of the Hall bar, non-local Hall voltage ($V_{nl}$) measured in the on-state of a monolayer $MoS_2$ device (40V < $V_g$ < 60V) can reach ~0.6V at T = 300K and increase to ~1.2V at T = 4K, as compared to ~10mV – 50mV $V_{nl}$ readings in the on-state of a multi-layer device (20V < $V_g$ < 40V), shown in Fig. 2(A, C). As mentioned above, VHE does not exist in multi-layer $MoS_2$ since inversion symmetry is not broken and transport does not occur in K and K' valleys. Therefore, the detected finite $V_{nl}$ signals in multi-layer devices can only be associated with Ohmic contribution or any other unknown effects. The magnitude of the non-local voltage due to the Ohmic contribution is expected to be dependent on the sheet resistance ($\rho_{sh}$) of the channel and device geometry: $V_{Ohmic} = I_{DC}\rho_{sh}\frac{W}{W_1}e^{\frac{-\pi L}{W}}$ (*21*), where L is the channel length and W and $W_1$ are the width of the channel and the current electrode, respectively (labeled in Fig. 1A). Using individual $I_{DC}$ and $\rho_{sh}$ measured for monolayer and multi-layer MoS$_2$, we are able to calculate the Ohmic contribution as a function of the back gate voltage ($V_g$) for each device, as presented in Fig. 2(B, D). We notice that the magnitude of the measured $V_{nl}$ of the multi-layer device from Fig. 2C matches the values of the calculated Ohmic contribution, while more than 1 order of magnitude larger $V_{nl}$ signals are measured in the monolayer MoS$_2$ device with an opposite temperature trend that we will discuss later. This significant magnitude difference in measured non-local voltages is also supported by a detailed potential analysis that resembles our experimental setup as shown in Fig. 3. Using experimentally measured contact resistance and MoS$_2$ sheet resistance obtained from the four-probe measurement, only a fraction of the supply voltage ($V_{ds}$ = 5V) is actually applied across the injector lead, i.e. $V_{in}$ = 1.8V. We then simulated in SPICE a resistor network with $4\times10^6$ identical resistors uniformly distributed over the Hall bar and observed that when a constant voltage of 1.8V is applied at the injector, the non-local voltage drop across the detector lead in the given geometry due to the Ohmic contribution is expected to be ~ 29 mV. This picture can get more complicated by the gate field controlled Schottky-barrier contacts (*22*). Nevertheless, we conclude that the magnitude of $V_{nl}$ due to the Ohmic contribution calculated from the resistor network is in good agreement with the experimental measurements in multi-layer MoS$_2$ devices. We benchmark the resistor network based SPICE simulation with the analytical equation in (*20*), section 2.1.

In addition to the magnitude of $V_{nl}$, its temperature dependence provides another evidence in support of the VHE being responsible for the non-local carrier transport in monolayer MoS$_2$. Fig. 4A shows increasing $V_{nl}$ with decreasing temperature down to 50K in monolayer MoS$_2$, while a completely opposite trend is observed for the multi-layer in Fig. 4B. Note that, since a voltage source, $V_{ds}$, is used in our measurements (instead of a constant current source), the temperature dependence of $V_{ohmic}$ due to the sheet resistance ($\rho_{sh}$) is expected to be cancelled out. However, finite contact resistance ($R_c$) needs to be considered in all MoS2 devices, which prevents $\rho_{sh}$ to be eliminated in the evaluation of the Ohmic contribution. In fact, it is expected that $V_{Ohmic} = \frac{V_{ds}}{\left(2R_C + \rho_{sh}\frac{L_1}{W_1}\right)} \rho_{sh} \frac{W}{W_1} e^{\frac{-\pi L}{W}}$. Different temperature dependence of $R_c$ and $\rho_{sh}$ are observed in four probe measurements, presented in (*20*), section 1.2. Indeed, the increasing $V_{nl}$ with increasing temperature observed in Fig. 4B for the multi-layer MoS2 device (dots) can be fitted by the modified $V_{ohmic}$ equation, considering the contribution from the contact resistance (lines). On the other hand, the increasing $V_{nl}$ with decreasing temperature down to 50K for the monolayer MoS2 device is expected for enhanced intervalley scattering length (λ) at low temperatures (*9, 23, 24*), confirming that valley transport is responsible for the observed large signals. More detailed analyses of non-local signals and inter-valley scattering length will be discussed in the following section.

Interestingly, this increasing $V_{nl}$ with decreasing temperature trend stops at T ~ 50K and reaches its maximum value. This unique maximum point results from two extreme limits of λ approaching either zero or infinity. While smaller $V_{nl}$ is expected with increasing temperature due to a shorter λ, large λ at temperatures lower than 50K can also lead to reduced non-local resistances, $R_{nl} = V_{nl}/I_{DC}$. This transition can actually be analogized to the well-studied quenched Hall effect (*25, 26*), where the Hall voltage vanishes when the carriers' longitudinal velocity is much higher than the transverse velocity. We suggest that the observed non-monotonic temperature dependence of $V_{nl}$ for monolayer MoS2 is an outcome of the monotonically increasing λ with decreasing temperature. We highlight that such temperature dependence of λ is consistent with the recent observation of increased intervalley scattering rate at higher temperatures in TMDs, which is attributed to phonon activated intervalley relaxation (*27*).

We will now quantitatively analyze this interesting temperature dependence of $R_{nl}$ for monolayer MoS2 using a self-consistent theoretical model describing the VHE. This model, similar to other theoretical descriptions in the literature (*18*, *21*) assumes a uniform and rectangular geometry without considering the arm lengths (Fig. 4C). Also following (*18*, *21*), we use a circuit model that is equivalent (*28*, *29*) to the standard spin-diffusion equation used in the context of materials with spin Hall effect to describe the VHE by defining the valley Hall angle as, $\theta = \sigma_{xy}/\sigma_{xx}$, where $\sigma_{xx}$ and $\sigma_{xy}$ denote longitudinal and transverse Hall conductivities, respectively. The valley Hall conductivity includes both intrinsic and extrinsic contributions and can be written as $\sigma_{xy} = \sigma_{xy}^{in} + \sigma_{xy}^{ex}$ (*30*). When the Fermi-level lies close to the conduction band minima, a condition that is fulfilled by our MoS2 devices (see (*20*), section 2.5), $\sigma_{xy}^{in}$ dominates over $\sigma_{xy}^{ex}$ (*30*). Using $\sigma_{xy}^{in} \sim \frac{2e^2}{h}$ (*16*) and measured $\sigma_{xx}$, we estimate $\theta \sim 0.4$ at T = 50K for our devices which is similar to Gorbachev et al.'s estimation (*16*) (see (*20*), section 2.5 for a detailed calculation of θ as a function of temperature). As also noted in (*18*), when θ is not small (i.e. θ ~ 1), one needs to self-consistently solve $R_{nl}$ considering the feedback impact of iVHE that behaves as a load to the generating section (induced by the direct VHE), and the impact of VHE that serves as a load to the detecting section (governed by the iVHE). Our circuit model automatically captures such self-consistencies to arbitrary order when solved in SPICE, but it is possible to derive an analytical equation considering only the iVHE at the generator side and the VHE at the detector side as second order effects. Further, our model takes the width of the arms explicitly and we can analytically obtain the following expression for the non-local resistance (see (*20*), section 2.2 for detailed derivation):

$$R_{nl} \equiv \frac{V_{nl}}{I_{DC}} = \frac{2\rho\lambda W exp\left[-\frac{L}{\lambda}\right]\sinh\left[\frac{W_1}{2\lambda}\right]\sinh\left[\frac{W_2}{2\lambda}\right]\theta^2}{\left(exp\left[\frac{W_1}{2\lambda}\right]W_1 + 2\lambda\sinh\left[\frac{W_1}{2\lambda}\right]\theta^2\right)\left(exp\left[\frac{W_2}{2\lambda}\right]W_2 + 2\lambda\sinh\left[\frac{W_2}{2\lambda}\right]\theta^2\right)} \quad (1)$$

where $\lambda$ is the intervalley scattering length, $W_{1,2}$ are the widths of the generating and detecting arm respectively, $\rho_{sh}$ is the sheet resistance and W is the width of the channel in Fig. 4C. We combine our VHE model (*29*) with non-magnetic circuit models that are also derived from a valley-diffusion equation (without any spin-orbit coupling) to obtain the infinite valley-loads on both ends, as well as to obtain the valley-diffusion in the middle channel whose length is denoted by L, based on the spin-circuit modeling described in (*28*). Conversely, the VHE model only considers charge transport in the vertical direction and valley coupled topological current in the longitudinal direction. It is important to note that, Eq. 1 is validated by a self-consistent numerical simulation of the composite valley-circuit in SPICE simulations and can be analytically reduced to the expression generally used in the literature (*21*), if we assume $\theta_2 \ll 1$ and $W_{1,2}/\lambda \ll 1$, yielding: $R_{nl} = \frac{1}{2}\left(\theta^2 \frac{W}{\sigma\lambda}\right)exp\left(\frac{-L}{\lambda}\right)$. It is clear from the complete (Eq. 1) and reduced equation that the two extreme limits of $\lambda$ naturally lead to an optimal intervalley scattering length to reach the maximum non-local resistance value.

This unique behavior enables us to quantitatively extract $\lambda$. Suggested by Eq. 1, the temperature dependence of $R_{nl}$ comes from that of $\lambda$ and $\theta$. With the calculated $\theta$ (T) shown in the inset of Fig. 4D ((*20*), section 2.5), we are able to fit the normalized non-local resistance, $R_{nl}^{norm}$ curve (dashed blue line in Fig. 4D, labeled as "Empirical") by tuning $\lambda$ (T). Since different physical mechanisms are responsible for the decreasing $R_{nl}^{norm}$ in the low and high temperature regimes, we can separately fit the high temperature trend with a power law function of $\lambda \propto T^{-0.73}$, which is in line with the temperature dependence of intervalley scattering that will be discussed later. Fitting for T > 75K regime is shown as the solid green line in Fig. 4D. λ (T) is then quantitatively extracted from $R_{nl}^{norm}$ and plotted (blue dots) in Fig. 4E, in a good agreement with the power law fitting at high temperatures. Furthermore, using the analytical expression in (*31*) describing both acoustic and optical intervalley phonon scattering together with the field effect mobility extracted from type II measurements, we are able to analytically derive λ (T) as shown as the solid line in the inset of Fig. 4E (see (*20*), section 2.7). A power law fitting of $\lambda \propto T^{-0.6}$ (dashed line) is obtained here, which is consistent with the experimental fitting of $\lambda \propto T^{-0.73}$ at T > 100K. At low temperatures, the extraction of $\lambda > 1\ um$ from the experimentally measured non-local signals is comparable to other valley Hall systems, as reported in (*16–18*). In general, λ is believed to be governed at low temperatures by atom-like defects that provide the necessary momentum required for carriers to scatter between K and K' valleys in the conduction band. In MoS$_2$, these atom-like defects arise due to molybdenum and sulfur vacancies. Recently, it has been pointed out that owing to the symmetry of atomic defects, only molybdenum vacancies can participate in intervalley scattering (*32*). Fourier transform scanning tunneling spectroscopy studies also provide further evidences (*33, 34*). The relatively large λ on the micron-scale extracted from our devices could be a result of relatively low molybdenum vacancy density in our MoS$_2$ sample.

Last, non-local signals measured with in-plane magnetic field applied up to 5T were presented in (*20*), section 2.8. As expected, no impact from the magnetic field is observed, indicating the robustness of the valley polarization in monolayer MoS$_2$ and further excluding the possibility of spin Hall effect responsible for our measurements. Therefore, these results once again resonate with the mechanism of the valley Hall effect (*5*, *35*).

In summary, we report electrically generating and detecting valley-coupled topological current in monolayer MoS$_2$ for the first time. Our approach provides a unique way to integrate charge, spin and valley degrees of freedom, which can be useful for emerging device technologies.

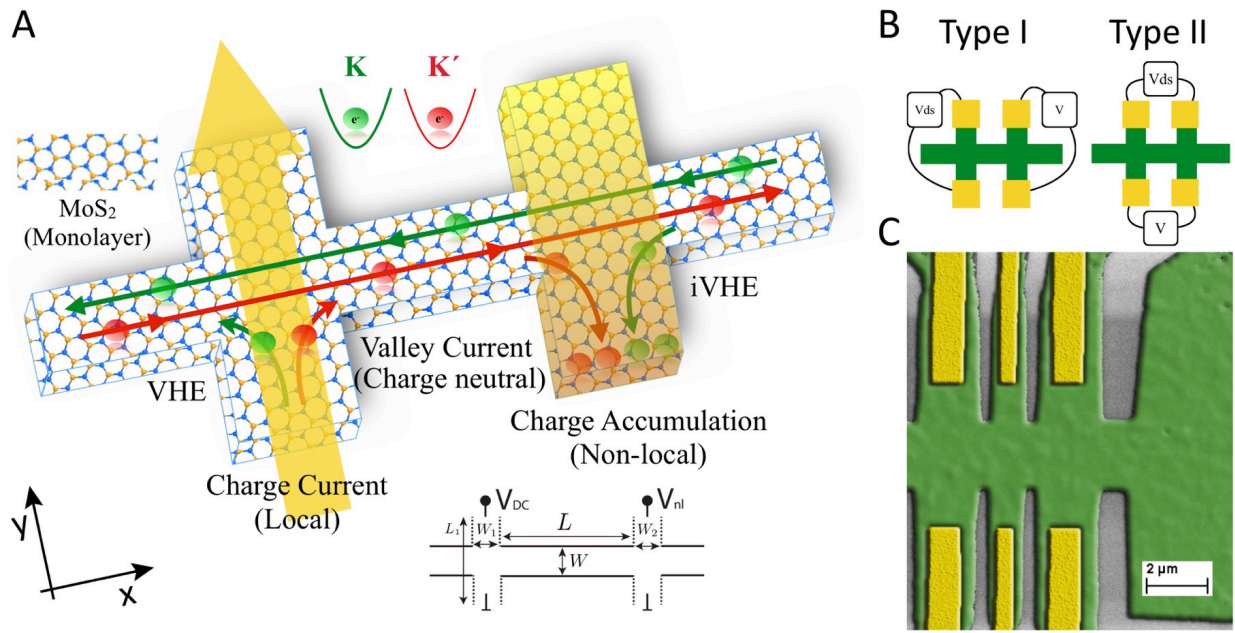

**Fig. 1. Valley coupled topological current** (A) Schematic of valley coupled topological current due to VHE and iVHE in monolayer MoS$_2$ and the device geometry (Bottom) where W$_1$ = 1 um, W = W$_2$ = 2 um, L$_1$ = 4.5 um, and L = 0.5 um. (B) Schematics of two measurement set-ups, type I and type II. (C) A patterned MoS$_2$ flake (green) and lithography defined metal electrodes (yellow).

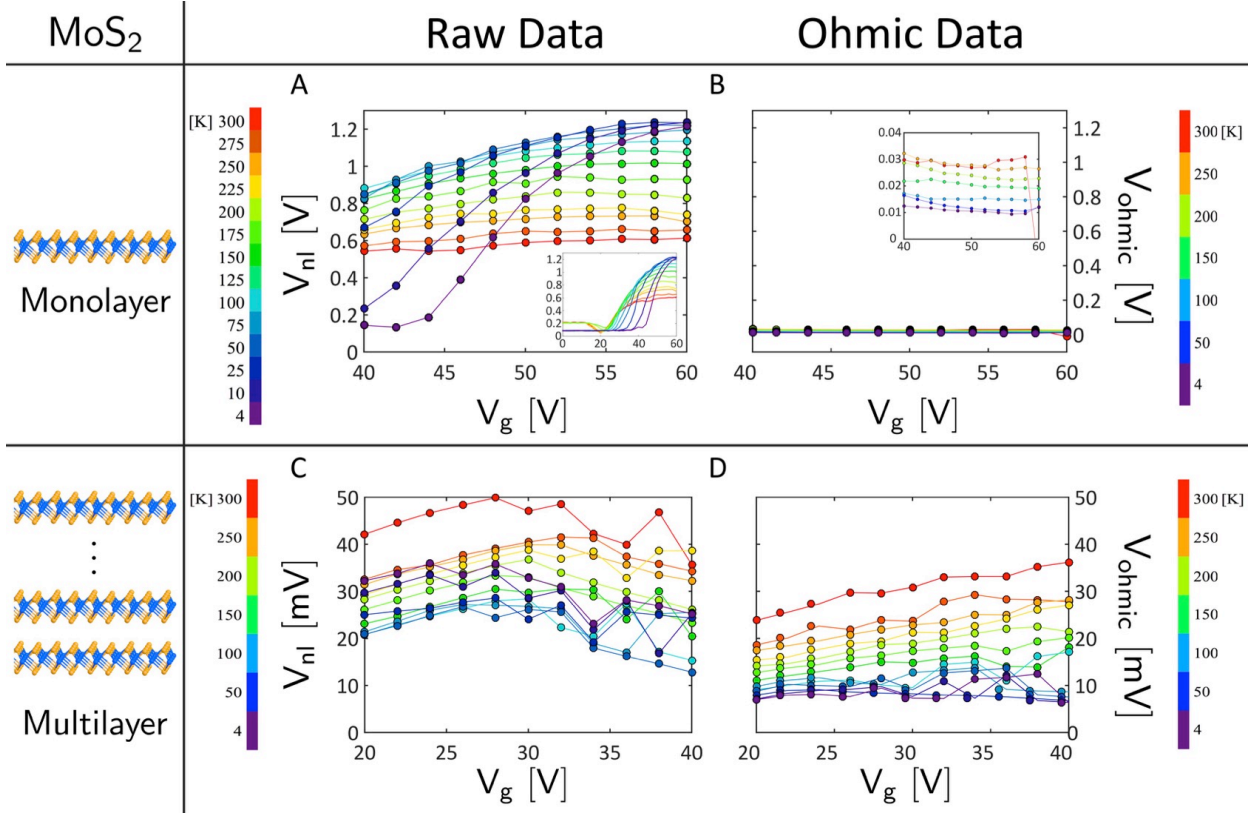

**Fig. 2. Comparison of non-local voltages obtained in monolayer and multilayer MoS$_2$ devices.** (A) Measured non-local voltage with respect to global back gate voltage V$_g$ in monolayer MoS$_2$ using type I setup. (Inset) Full range of V$_g$. Note that, data points in the range of V$_g$ < 40V are not included in analysis since these large device resistances become comparable to the input impedance of the nano-voltmeter. (B) Ohmic contribution calculated from the measured sheet resistance: $V_{Ohmic} = I_{DC} \rho_{sh} \frac{W}{W_1} e^{\frac{-\pi L}{W}}$ as a function of V$_g$, plotted with the same y-axis range as in (A). (Inset) shows the zoom-in data. (C-D) Non-local voltage response in a multilayer MoS$_2$ device for the same measurements performed in (A-B). Note that, the y-axis in both plots has a unit of mV.

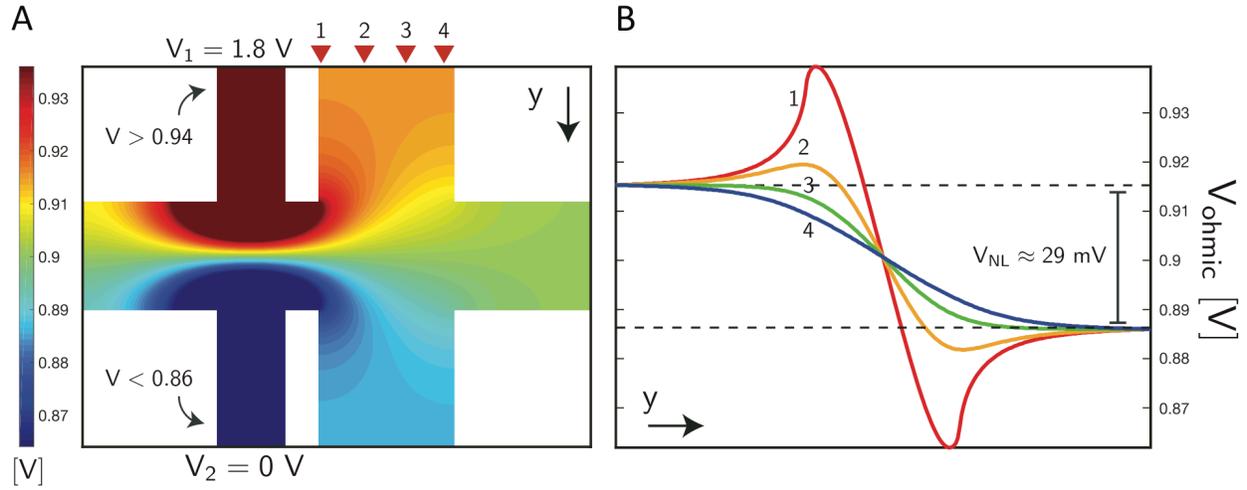

**Fig. 3. Electric potential mapping from a SPICE-based resistor network simulation.** (A) SPICE simulation of a resistor grid with ~ $4 \times 10^6$ uniform resistors where each resistor corresponds to ~ 3 nm channel length, with (x=1500, y=1400) points. $V_{ds}$ applied at the two ends of the injector are $V_1$ = 1.8V and $V_2$ = 0 V, respectively. Values greater than 0.94 V and less than 0.86 V are denoted with the same colors to resolve the non-local voltage distribution. (B) Voltage profiles along the y direction for 4 different positions denoted by arrows (1-4) in (A). Non-local voltage difference under open circuit condition is calculated to be ~ 29 mV.

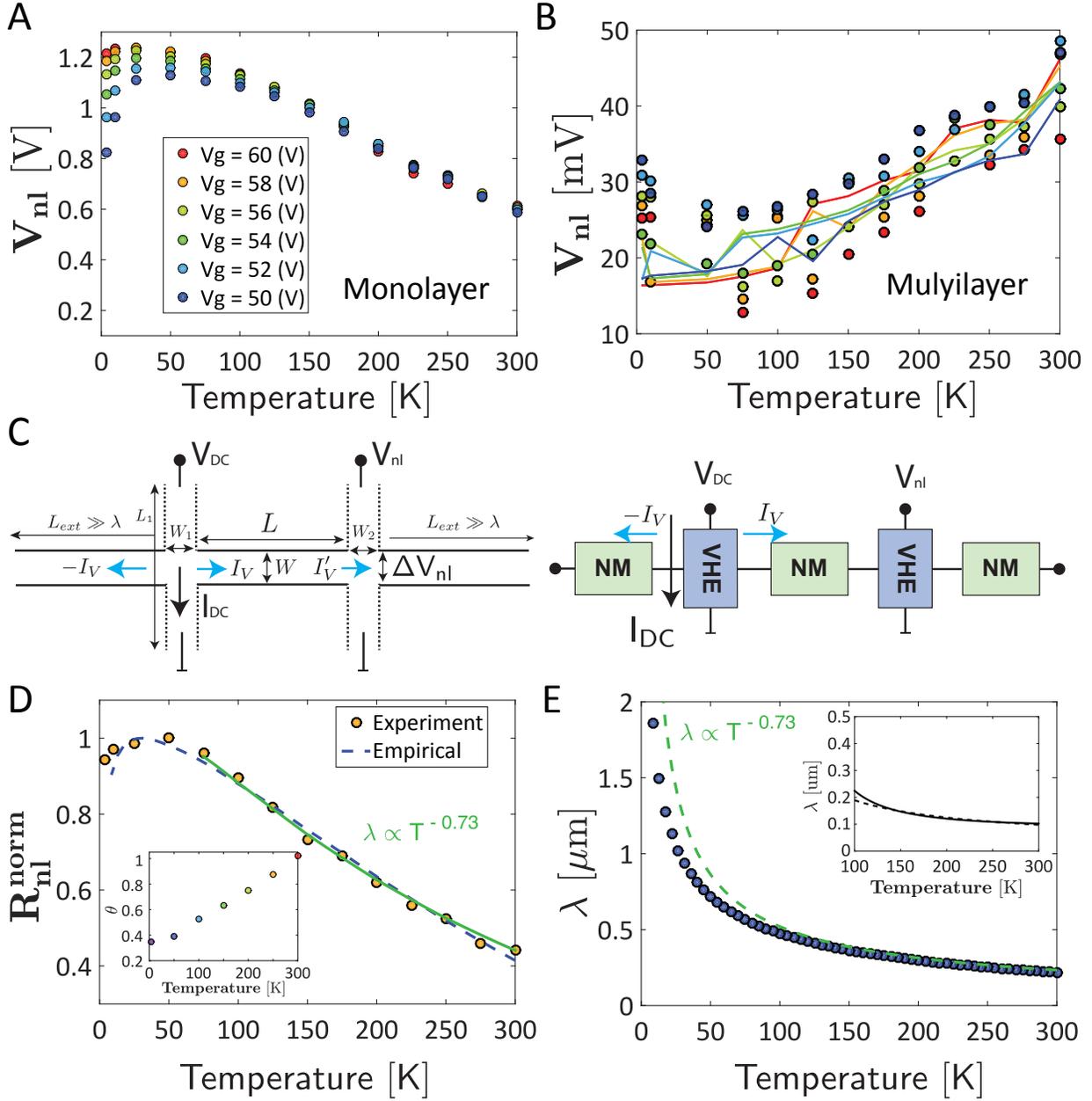

**Fig. 4. Temperature dependence and extraction of intervalley scattering length.** (A) Measured $V_{nl}$ as a function of temperature at different $V_g$ for monolayer $MoS_2$. (B) Temperature dependence of multilayer $MoS_2$ at different $V_g$ (dots) and the calculated trends (lines) using the modified Ohmic equation, $V_{Ohmic} = \frac{V_{ds}}{\left(2R_C + \rho_{sh}\frac{L_1}{W_1}\right)} \rho_{sh} \frac{W}{W_1} e^{\frac{-\pi L}{W}}$, with the consideration of the contact resistance contribution (see (20), section 1.2). Note the trends of $V_{nl}$ with respect to temperature in (A) and (B) are completely opposite. (C) Device geometry and corresponding valley-circuit model that define the geometric parameters in Eq. 1. Details are given in (*20*), section 2.2. (D) Temperature dependence of $R_{nl}^{norm}$ (normalized to the maximum point) measured at $V_g$ = 58 V (orange dots in Fig. 4A). The empirical fittings use $\lambda(T) = 5.5\,T^{-0.47} - 0.16$ (dashed blue line) and $\lambda = 15\,T^{-0.73}$ at T > 100 K (green line). (Inset) Calculated temperature dependence of valley Hall angle, θ. (E) λ (T) extracted from $R_{nl}$ and the power law dependence described in (D). (Inset) Theoretically calculated intervalley scattering length (solid) and $\lambda \propto T^{-0.6}$ to guide the eye (dashed).

**Acknowledgments:**

We would like to acknowledge fruitful discussions with Prof. Joerg Appenzeller and Prof. Supriyo Datta. P. U. would like to thank Héctor Ochoa for useful discussions on intervalley scattering in TMDs. K. Y. C. acknowledges support from ASCENT, one of six centers in JUMP, a Semiconductor Research Corporation (SRC) program sponsored by DARPA. T. H., S. Z., and Z. C. gratefully acknowledge the support of this work by the Semiconductor Research Corporation (SRC)'s Nanoelectronic Computing Research (nCORE) program through the NEW LIMITS center.

# Supplementary Materials for

## Direct Observation of Valley Coupled Topological Current in MoS$_2$

Terry Y.T. Hung[1,2], Kerem Y. Camsari[1], Shengjiao Zhang[1,2], Pramey Upadhyaya[1], and Zhihong Chen[1,2]*

Correspondence to: zhchen@purdue.edu

**This PDF file includes:**

    Materials and Methods
    Supplementary Text
    Figs. S1 to S10

# 1 Materials and Methods

## 1.1 Device fabrication

CVD grown $MoS_2$ films were transferred to 90nm $SiO_2$ substrates with highly doped Si on the back side serving as a global back gate ($V_g$). The transfer process includes: 1) the sample was spin-coated with Polystyrene (PS) followed by immersing in DI water; 2) the PS/$MoS_2$ stack was then detached from the substrate and scooped up by the receiving $SiO_2$ substrate; 3) PS was subsequently dissolved by toluene and bathed in acetone and isopropyl alcohol (IPA) to thoroughly clean it. Standard e-beam lithography using PMMA A4 950 resist was employed to pattern electric contacts on the CVD $MoS_2$ flakes. Ti/Au (20/80nm) was deposited in an e-beam evaporator followed by a lift-off process in acetone. CVD grown BN film was transferred from Cu foil onto the devices through a process that involves etching the Cu foil with iron chloride ($FeCl_3$) and immersing it in diluted HCl and DI water alternatingly for few times before scooping up. This BN layer was inserted to minimize device degradation from PMMA residues after the RIE etching process. RIE etching mask was defined by e-beam lithography using PMMA A4 950 resist and BN/$MoS_2$ flakes were etched using Ar/$SF_6$ for 10 seconds. The final devices were annealed in forming gas ($N_2$/$H_2$) at 300°C for three hours followed by vacuum annealing ($\sim 10^{-8}$ torr) at 250°C for 4 hours to minimize PMMA residue and threshold voltage shift due to trap charges.

## 1.2 Optical and electrical characterization

Raman and contrast in optical images were used to confirm the thickness of the $MoS_2$ flakes in both of the VHE device (monolayer) and the control sample (multilayer) shown in Fig. S1(A, B). Raman spectra were obtained using an excitation wavelength of 532nm with a 50X objective lens. A Raman shift of 18 cm$^{-1}$ between the E$^1_{2g}$ and A$_{1g}$ modes in the monolayer is clearly different from that in the multilayer, as shown in Fig. S1A (*1–3*). Due to the measurement limitation in the current meter used in our experiments, the lowest current that can be measured was ~10$^{-10}$A (Fig. S1C), In MoS$_2$ devices, sub-threshold current above 10$^{-10}$A is dominated by tunneling current injected through the source/drain Schottky barriers, which shows weak temperature dependence. The observed threshold voltage shift is as expected since a larger gate voltage is required to compensate fewer carriers in the Fermi distribution at a lower temperature. Shown in Fig. S1(D, F), conventional four-probe measurements (type II) were used to extract sheet resistance (ρ) and contact resistances (R$_c$) for monolayer and multilayer devices, respectively (*4, 5*).

# 2 Supplementary Text

## 2.1 Details of resistor network for Ohmic contribution

In order to simulate the Ohmic contribution in the multilayer sample, we constructed a general SPICE network that matches the known analytical results for extremely simple geometries as shown in Fig. S2A. Our resistor network however can be "patterned" to arbitrary shapes and structures by placing very large resistor values to patterned regions (as shown in Fig. 3A of the main text).

## 2.2 Derivation of non-local resistance, R$_{nl}$

In this section, we outline the derivation details of Eq. 1, starting from a lumped "valley-circuit" model whose results are equivalent to those of the commonly used spin-diffusion equations (*6*). We then compare the analytical expression with a fully self-consistent SPICE-based numerical solution of the circuit.

Fig. S3 shows the circuit diagram that is based on (*7*). The lumped model combines non-magnetic (NM) regions that act as boundary conditions that are much longer than the diffusion length (λ) with two VHE layers that are bridged by another NM region that the valley polarized carriers diffuse over. We neglect the VHE physics in this middle layer but explicitly consider the spin-diffusion and loss. The VHE layers are composed of a charge-circuit and a valley-circuit that treat the charge and spin flows differently, as in (*8*). The model takes into account both the direct VHE and the inverse VHE with dependent current sources in the valley-circuit $I_1$, $I_2$ and in the charge circuit $I_3$, $I_4$, respectively. Therefore, the model captures effects such as self-induced inverse VHE due to a charge current flowing in the injection layer and a self-induced direct VHE in the detection layer due to an induced open-circuit voltage.

We define σ as the sheet conductivity of the material (σ = $σ_{xx}$t) where $σ_{xx}$ is the longitudinal conductivity and t is the thickness of the sample. The charge and valley conductance are defined in Fig. S3. We assume that a constant charge current $I_{DC}$ is being injected between nodes $V_{1c}$ and $V_{2c}$ and this gives rise to an open-circuit, non-local voltage $\Delta V_{NL}$ between nodes $V_{5c}$ and $V_{6c}$. We are then interested in a closed-form expression relating these two quantities, $R_{nl} \equiv \Delta V_{NL}/I_{DC}$. We consider three terms contributing to this expression:

- $i_1$: Self-generated VHE current (opposing) due to an injected current $I_{DC}$.
- $i_{2R}$: Direct VHE current due to an injected $I_{DC}$.
- $i_{2L}$: Direct VHE current (opposing) due to an induced $\Delta V_{NL}$.

We ignore the higher order terms assuming they get progressively smaller since θ < 1, and later show (Fig. S4) that the results are in good agreement with a full SPICE-based solution of the circuit without any assumptions. We start with the derivation of the current $i_1$ which increases the effective resistance of the injecting layer, similar to the Spin Hall Magnetoresistance effect. With a straightforward solution of the circuit we find:

$$i_1 = \frac{2V_{DC}\sigma\theta}{1 + exp[W_1/\lambda]} \qquad (1)$$

Using this current term, we can specify the induced charge voltage (due to inverse VHE through the current source $I_3$) and solve for the modified $V_{DC}$ that develops under a constant injected current $I_{DC}$:

$$V_{DC} = \frac{I_{DC}W}{\sigma(W_1 + \theta^2\lambda(1 - exp[-W_1/\lambda]))} \qquad (2)$$

which, in the limit $\lambda \ll W_1$ reduces to, $R' = V_{DC}/I_{DC} = W/W_1\sigma(1 + \theta^2)$, implying that the resistance of the injector arm increased by a factor proportional to $\theta^2$ due to the self-induced inverse VHE. We then use Eq. (2) to derive the term $i_{2R}$.

$$i_{2R} = \theta\sigma V_{DC} \frac{exp[-(W_1 + L)/\lambda](exp[W_1/\lambda] - 1)}{exp[W_2/\lambda] + 1} \qquad (3)$$

Similarly, we obtain the current $i_{2L}$ by keeping $\Delta V_{NL}$ as a variable and combine it with Eq. (3) to self-consistently solve for a $\Delta V_{NL}$ in terms of $I_{DC}$. With full simplifications, we obtain the following expression:

$$R_{nl} = \frac{2\rho\lambda W exp\left[-\frac{L}{\lambda}\right] \sinh\left[\frac{W_1}{2\lambda}\right] \sinh\left[\frac{W_2}{2\lambda}\right] \theta^2}{\left(exp\left[\frac{W_1}{2\lambda}\right] W_1 + 2\lambda \sinh\left[\frac{W_1}{2\lambda}\right] \theta^2\right)\left(exp\left[\frac{W_2}{2\lambda}\right] W_2 + 2\lambda \sinh\left[\frac{W_2}{2\lambda}\right] \theta^2\right)} \qquad (4)$$

We note that this expression reduces to the well-known non-local resistance formula [4] under the following limits, $\theta^2 \ll 1$ and $W_{1,2}/\lambda \ll 1$, yielding:

$$R_{NL} = \frac{1}{2}\left(\theta^2 \frac{W}{\sigma\lambda}\right) exp\left(\frac{-L}{\lambda}\right) \tag{5}$$

### 2.3 Additional non-local measurements for multilayer MoS$_2$ devices

Experimentally, we performed the same non-local measurements for multilayer MoS$_2$ devices in which the channel is known to have inversion symmetry and carrier transport is not governed by K and K' valleys, so the measured non-local voltage is attributed to the Ohmic contribution only. Fig. S5 shows qualitative agreement between the measured local V$_{nl}$ in multilayer MoS$_2$ devices and the Ohmic contribution equation $V_{Ohmic} = I_{DC}\rho_{sh}\frac{W}{W_1}e^{\frac{-\pi L}{W}}$ (9) for different channel lengths.

### 2.4 Additional non-local measurements from other monolayer MoS$_2$ devices

In Fig. S6, we present additional room temperature data from other monolayer devices to show the reproducibility of the VHE. All of them, including the one presented in the main text, show one to two orders of magnitude larger non-local signals than the Ohmic contribution signals under the same measurement conditions. We attribute the magnitude difference to the device-to-device fabrication variations.

### 2.5 Detailed θ calculation and its temperature trend

In the main text, we define $\theta = \frac{\sigma_{xy}}{\sigma_{xx}}$. In order to estimate $\theta$, we calculate $\sigma_{xy}$ due to intrinsic (Berry's phase) contribution to the valley Hall conductivity, $\sigma_{xy}^{in}$, while directly extracting $\sigma_{xx}$ from type II measurements shown in Fig. S1D. It should be noted that in doing so, we have ignored possible impurity scattering-induced extrinsic contributions (*10, 11*) to the valley Hall effect. As pointed out in (*10, 11*), this approximation is justified for the case when Fermi-level ($E_F$) lies close to the conduction band edge. This is indeed the case for the voltage range explored in our measurements. In particular, $E(k) = \pm\sqrt{\Delta^2 + v^2\hbar^2 k^2}$, for $V_g$-$V_{th}$ = 40 V, the position of the Fermi-level (as measured from the middle of the band gap) is given by $E_F \sim \frac{n\pi\hbar^2}{2m_e^*} + \Delta \sim 0.89\ eV$. Here, $2\Delta \sim 1.72\ eV$ is the band gap for MoS$_2$, $\hbar$ is the Plank constant divided by $2\pi$, $m_e^* = 0.4 m_e$ is the electron effective mass in MoS$_2$, and $n \sim (V_g - V_{th})\epsilon_r\epsilon_0/t_{SiO_2}e$ is the surface charge accumulated by the gate, with $V_g - V_{th}$ being the overdrive voltage, $\epsilon_r\epsilon_0$ as the permittivity in SiO$_2$, and $t_{SiO_2} = 90\ (nm)$ as the dielectric thickness. The intrinsic valley Hall conductivity is given by (*12*)

$$\sigma_{xy}(E_F) = \sum_{\tau_z} \tau_z \sum_{s_z,\alpha} \frac{e^2}{\hbar} \frac{1}{(2\pi)^2} \int dk_x dk_y \Omega(k, \tau_z, \alpha) f(E_{k,\alpha}), \tag{6}$$

where $E_F$ is the Fermi-level, $\tau_z$ is the valley index ($\tau_z = -1$ for K and $\tau_z = +1$ for K'), $\alpha$ is the band index ($\alpha = -1$ for the valance band and $\alpha = +1$ for the conduction band), $s_z$ is the spin index ($s_z = -1$ for up spin and $s_z = +1$ for down spin), and $\Omega = \tau_z \frac{\Delta^2 v^2 \hbar^2}{2(\Delta^2 + v^2\hbar^2 k^2)^{3/2}}$ where $v^2 = \Delta/m_e^*$ (*10–12*). Here, putting $E_F \sim 0.89\ eV$ we find $\sigma_{xy} \sim \frac{2e^2}{h}$. This value is consistent with the fact that for the Fermi-level position close to conduction band minima, the valley Hall conductance is dominated by the filled valence bands. Substituting this value in the definition of $\theta$ and using $\sigma_{xx}$ from Fig. S1D, we plot $\theta$ v.s. temperature in the inset in Fig. 4D. In general, increasing the temperature decreases $\sigma_{xy}$. This is because conduction and valence band contribute opposite signs to $\sigma_{xy}$, and increasing temperature increases conduction band occupation at the expense of the valence band's population. However, we highlight that for the Fermi-level position near the conduction band minima, and the band gap of $1.72\ eV \gg k_B T$ for T = 300K, the value of $\sigma_{xy}$ is independent of temperature (as verified by directly calculating $\sigma_{xy}$ for T = 1K and T = 300K using Eq. 6 and noting a decrease of less than 0.4%). In this case, the temperature dependence of $\theta$ comes primarily from the temperature dependence of $\sigma_{xx}$ in Fig. S1D. In Fig. S7, we plot this temperature dependence of $\theta$, which is also presented in the inset of Fig. 4D in the main text.

## 2.6 Non-local internal resistance measurements

We add external resistors into the measurement set-up to extract the internal resistance ($R_{MoS2}$ = 2$R_c$ + 2$R_{arm}$ + $R_{cross}$) in the non-local arm in both monolayer and multilayer MoS$_2$ devices, as depicted in Fig. S8A. The measured voltage drop across the external resistor ($R_{ext}$), can be described by $V_{ext} = I_{ext} (R_{ext} R_{MoS2}) / (R_{ext} + R_{MoS2})$. Simpler expression can be derived by normalizing to its maximum point:

$$V_{norm} = \frac{R_{ext}}{R_{ext} + R_{MoS2}} \qquad (7)$$

By changing over a large range ($10^2$ to $10^8$ Ω) of external resistance values ($R_{ext}$) depicted in Fig. S8A and fitting with Eq. 7, we are able to extract the internal resistance. We notice that $V_{nl} \neq \Delta V_{nl}$, since $\Delta V_{nl}$ should be a fraction of $V_{nl}$, denoted in Fig. 4C in the main text. Intuitively, one might think that the ratio of $\Delta V_{nl}$ to $V_{nl}$ should be equal to the ratio of $R_{cross}$ to $R_{MoS2}$ (non-local total resistance) shown in the Fig. S8A. However the extracted internal resistance (24MΩ) by fitting presented in Fig. S8B dose not agree with the non-local total resistance (7MΩ) in monolayer MoS$_2$. In contrast, the extracted internal resistance (25kΩ) presented in Fig. S8C is very close to the non-local total resistance (35kΩ) in multilayer MoS$_2$. Furthermore, we use SPICE resistor network discussed in section 2.1 to simulate this internal resistance extraction for multilayer MoS$_2$ with two vastly different resistor values of $10^3$ (red) and $10^6$ (blue) shown in Fig. S8D. As expected, it shows very good agreement between the extracted internal resistance and the non-local total resistance. Thus we speculate that for VHE governed monolayer MoS2 devices, it is not sufficient to take the resistance ratio ($R_{cross}$ to $R_{MoS2}$) for the internal resistance calculation. Instead, one should probably carefully take into account some resistance amplification due to the VHE over the entire electrode lead. Both Eq. 1 in the main text and SPICE capture the physics of the (L × W) rectangle shown in Fig. 4C without considering the extended arm. Further experiments, such as varying the arm length, direct measuring $R_{cross}$ and independently controlling the contact and channel resistance, are required to understand the discrepancy of internal resistances between the valley Hall and non-valley Hall systems.

## 2.7 Detailed λ calculation and its temperature trend

Within the deformation potential approximation, the analytical expression of intervalley scattering rate τ, as obtained from Fermi's golden rule, is given by (13),

$$\frac{1}{\tau} = g_d \frac{m^* D_0^2}{2\hbar^2 \rho \omega} [N\Delta_1 + (N+1)\Delta_2] \tag{8}$$

Here $g_d$ is the valley degeneracy for the final electron states, $m^*$ is density-of-state effective mass for the K valley, $D_0$ is the deformation potentials in K valley ($D_0^{op}, D_0^{ac}$ are for optical and acoustic phonon respectively), $\rho$ is the mass density ($= 3.1 \times 10^{-7}\ g/cm^2$) for MoS2, $\hbar\omega$ is phonon energy, $N$ is Bose-Einstein distribution and $\Delta_1, \Delta_2$ are the onset of scattering for phonon absorption and emission respectively. Using Eq. 8, $\lambda = \sqrt{D_{diff}\tau}$, Einstein relation for diffusion coefficient ($D_{diff} = \mu k_B T/q$), and experimentally extracted field effect mobility ($\mu$), we calculate intervalley scattering length (λ) in high temperature regime (T > 100K) shown in Fig. S9. The calculated λ (T) (solid line) can be fitted with a power law dependence of $\lambda \propto T^{-0.6}$ (dashed line).

## 2.8 Applied in-plane magnetic field

The observation of robustness of valley polarization under in-plane magnetic field up to 5T mentioned in the main text is shown in Fig. S10.

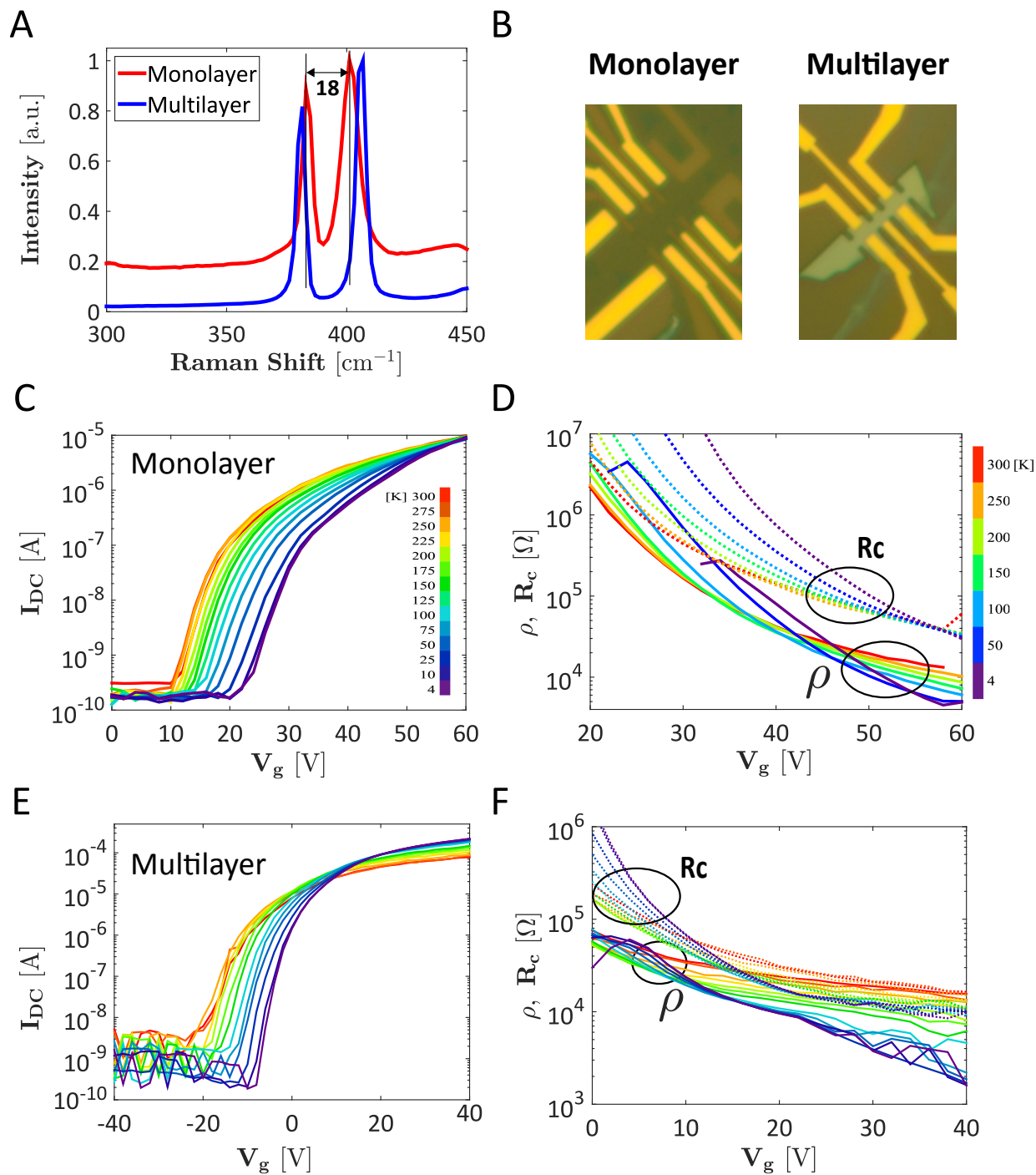

**Fig. S1. Device characterization.** (A) Two prominent Raman characteristic peaks for MoS$_2$ flakes. Monolayer presents a distinguished Raman shift of 18 cm$^{-1}$ between the $E'$ and $A'_1$ peaks. (B) Representative optical images for monolayer and multilayer MoS$_2$ devices. Transfer characteristics at different temperatures for monolayer (C) and multilayer (E) MoS$_2$ devices.

Four-probe measurements using type II set up described in the main text to extract sheet resistance $\rho$ and contact resistance $R_c$ in monolayer (D) and multilayer (F) devices.

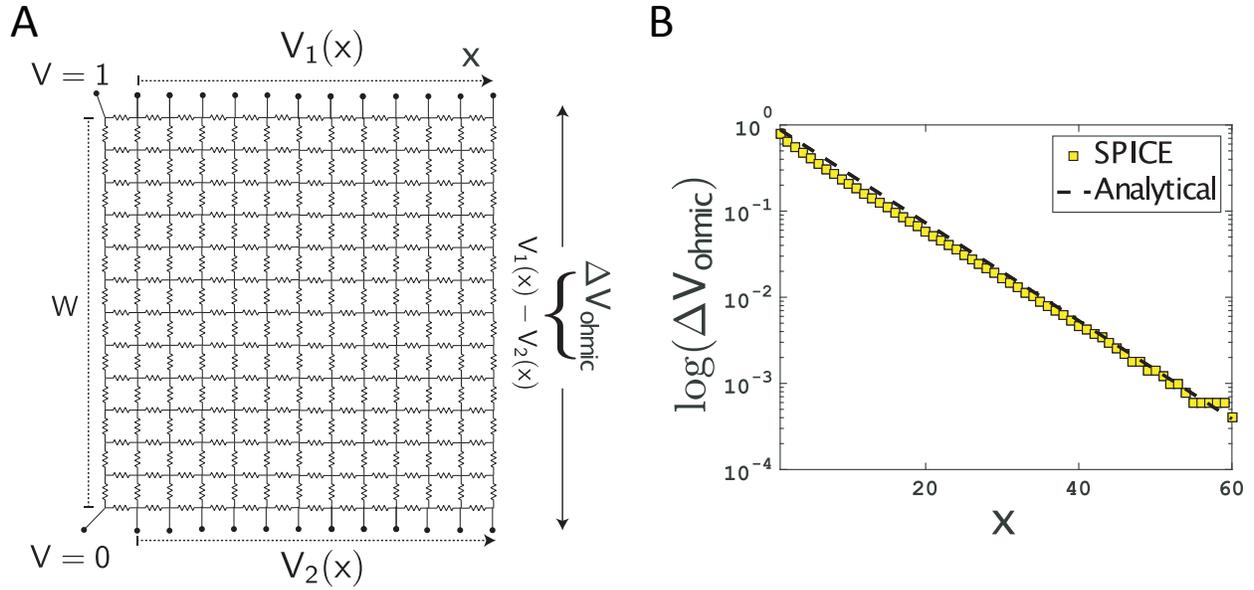

**Fig. S2. SPICE-based resistor network.** (A) A uniform resistor network (not to scale) to simulate the Ohmic contribution in SPICE. In a type I setup, a voltage of 1V is applied between top and bottom contacts, and the non-local voltage is measured as a function of the length in the network. (B) An approximately infinitely narrow strip (L/W ≈ 1000) with a resistor network containing more than a million resistors is simulated to compare with the known analytical formula $\Delta V_{ohmic} \propto \exp(-\pi x/W)$. SPICE-based results are in excellent agreement with the approximate formula. The SPICE-based network however can simulate arbitrarily patterned shaped and non-uniform structures as shown in the main text.

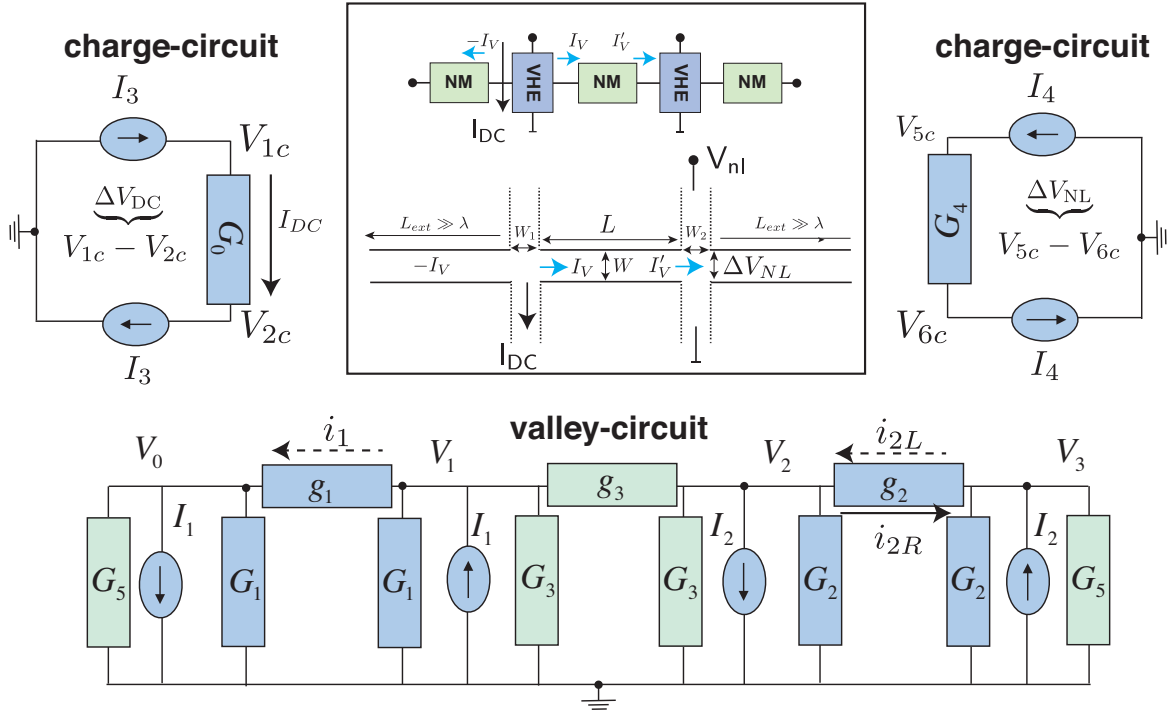

**Fig. S3. The lumped valley-circuit model that is used to derive Eq. 1 in the main text.** The charge-circuit captures the injected and induced charge currents and voltages in the vertical direction, while the valley-circuit captures the valley diffusion currents in the horizontal direction. The charge-circuit parameters are defined as: $G_0 = \sigma W_1/W$, $G_4 = \sigma W_2/W$, $I_3 = \sigma\theta(V_1 - V_0)$ and $I_4 = \sigma\theta(V_2 - V_3)$ where $\theta$ is the valley Hall angle, $\sigma$ is the sheet conductivity, and $W_1$, $W_2$, $W$ are the width of the injector, detector and the middle region, as shown in the figure. $(V_1 - V_0)$ and $(V_2 - V_3)$ are the non-equilibrium valley potentials that control the inverse valley Hall terms in the charge circuit. The valley-circuit parameters are defined as: $g_i = \sigma W/\lambda \, \text{csch}(W_i/\lambda)$, $G_i = \sigma W/\lambda \, \tanh(W_i/(2\lambda))$, where $i \in \{1, 2, 3\}$ with $W_3 = L$ and $G_5 = \sigma W/\lambda$. Finally, the current sources $I_1 = \theta\sigma V_{DC}$ and $I_2 = \theta\sigma V_{nl}$ where $V_{DC}$ is the applied voltage and $\Delta V_{NL}$ is the induced non-local voltage as defined in the figure. See text for the description of the current terms $i_{2R}$, $i_{2L}$ and $i_1$ that are used in the derivation.

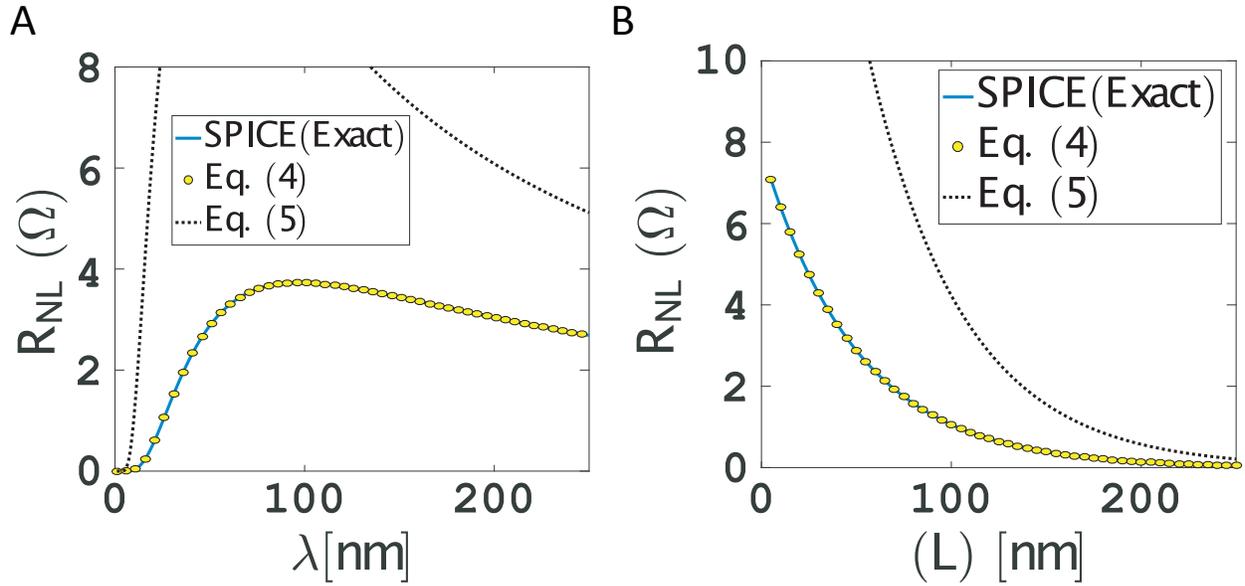

**Fig. S4. Comparison of analytical equations for $R_{nl}$ with the full SPICE simulation of the circuit shown in Fig. S3.** Excellent agreement is observed between SPICE and Eq. 4, while deviations are clearly shown for the reduced Eq. 5. The parameters are $\theta = 0.5$, $\sigma = 2$ mS, $W_1 = 50$ nm, $W_2 = 75$ nm, $W = 25$ nm, $L = 50$ nm for (A) and $\lambda = 50$ nm for (B). It is interesting to note that the expression based on Eq. 5 overestimates the magnitude of the signal and for large $\theta$ and a self-consistent model as described here (also as in [10]) is required.

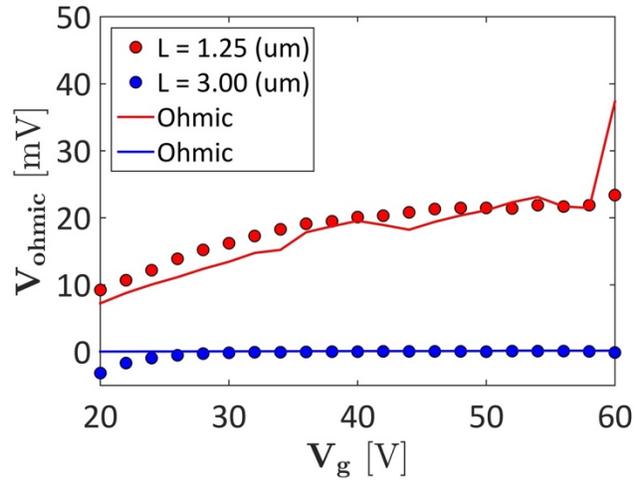

**Fig. S5. Long channel multi-layer MoS$_2$ devices.** Type I, non-local measurements were performed in additional multilayer MoS$_2$ devices with different channel lengths. Dotted lines are the calculated Ohmic contribution as described in the text. Close to zero non-local voltages are measured in the device with channel length of L = 3μm.

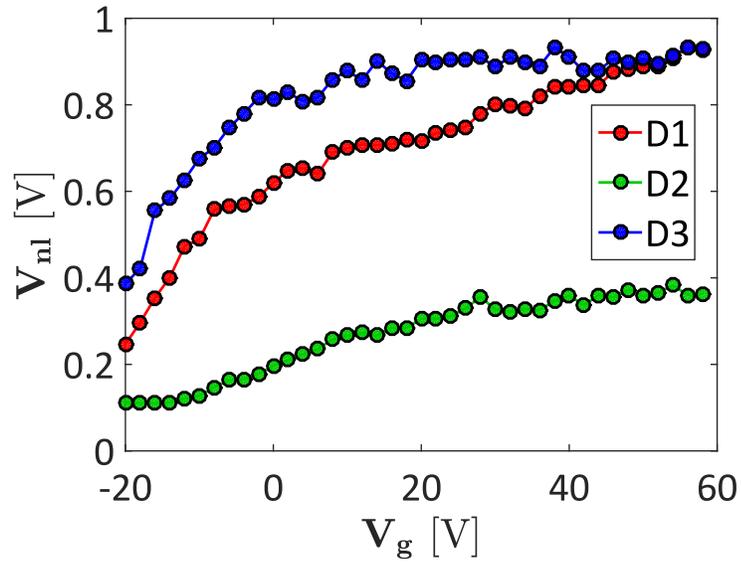

**Fig. S6. Additional monolayer MoS$_2$ device measurements.** Type II, non-local measurements for different monolayer MoS$_2$ devices to show the reproducibility and robustness of the VHE. They all have $W_1$ = 1 um, W = $W_2$ = 2 um, $L_1$ = 4.5 um, and L = 0.5 um as depicted in Fig. 1A.

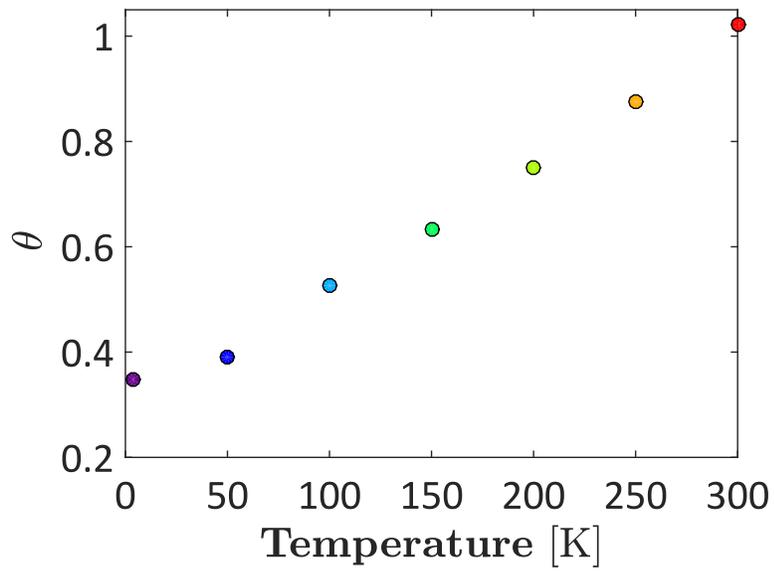

**Fig. S7. Temperature dependence of valley Hall angle.**

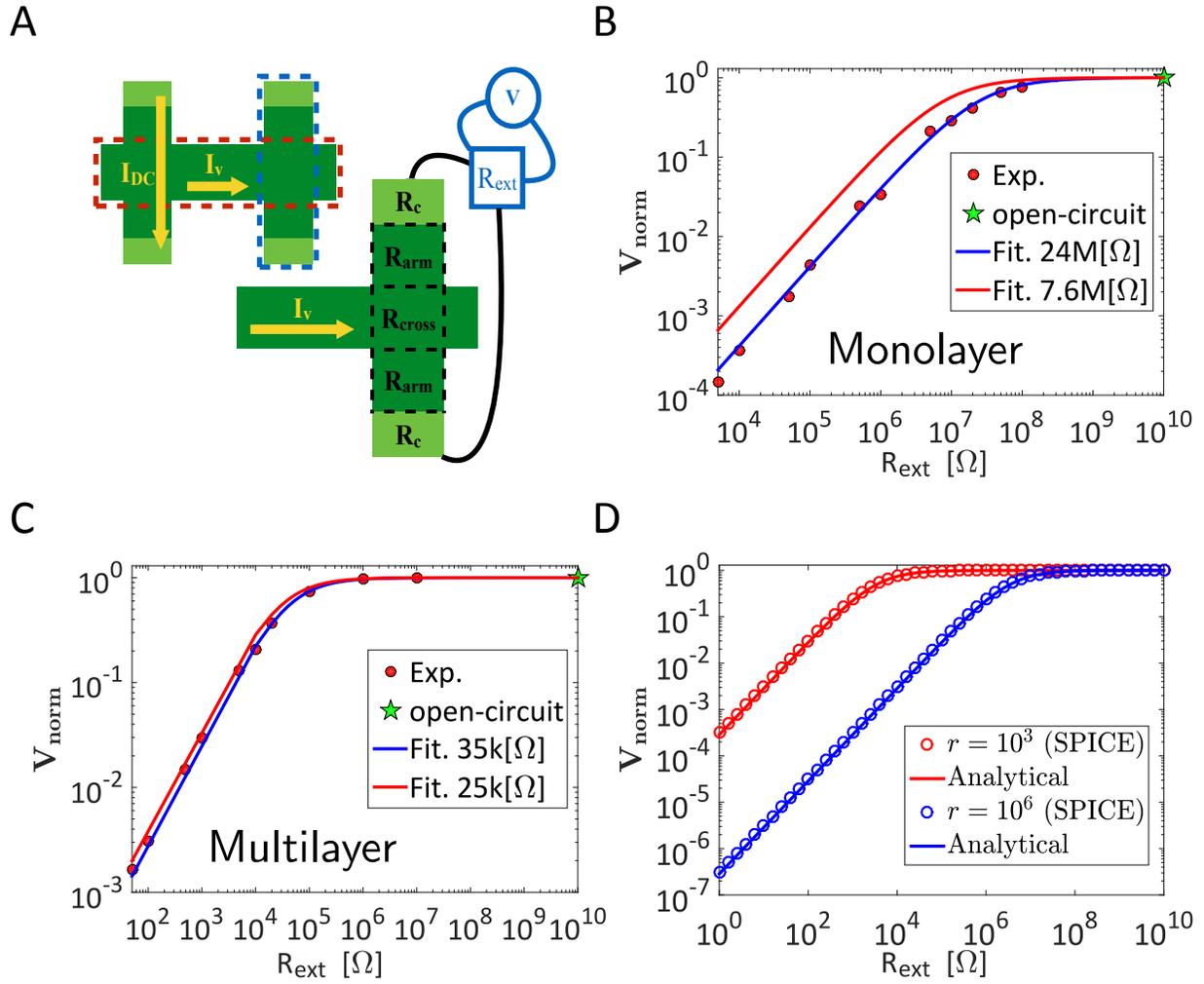

**Fig. S8. Extraction of internal resistance in the non-local electrode.** (A) Schematic of the measurement set-up with an external resistor. (B) Monolayer, (C) multilayer internal resistance ($R_{MoS_2}$) extraction and comparison between total resistance ($R_{tot}$) and internal resistance ($R_{MoS_2}$). (D) SPICE modeling in a uniform resistor Hall structure shown in Fig. S2.

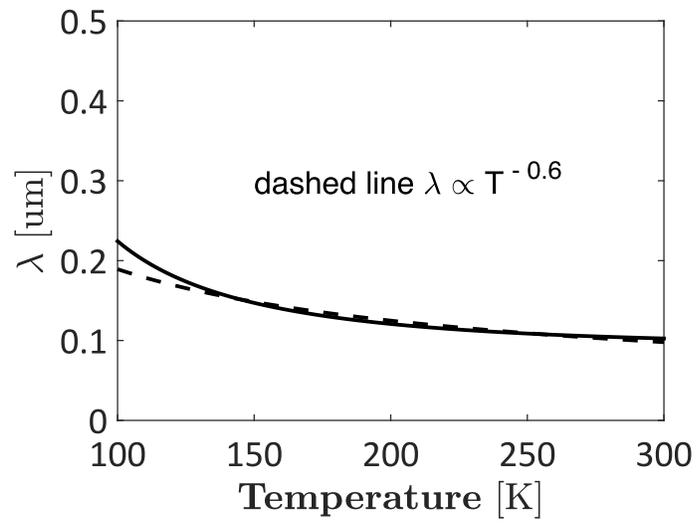

**Fig. S9. Temperature dependence of intervalley scattering length.**

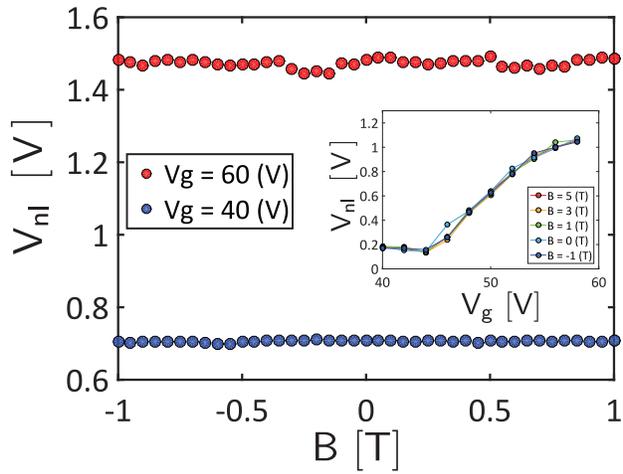

**Fig. S10. $V_{nl}$ measurements with in-plane magnetic field applied.** Measured $V_{nl}$ as a function of the applied in-plane magnetic field. (Inset) Non-local voltage, $V_{nl}$ as a function of $V_g$ under in-plane magnetic fields up to 5T.